\documentclass[12pt]{article}
\usepackage{a4wide,epsfig}

\def\eg{{\em e.g.}}

\def\ie{{\em i.e.}}
\def\etal{{\em et al.}}

\newcommand{\be}{\begin{equation}}
\newcommand{\ee}{\end{equation}}
\newcommand{\ba}{\begin{eqnarray}}
\newcommand{\ea}{\end{eqnarray}}

\begin{document}

\thispagestyle{empty}

\begin{center}

\vskip \baselineskip
{\Large\bf Dileptons and Medium Effects in Heavy-Ion Collisions} \\
\vskip 3\baselineskip

R.\ Rapp\,\footnote{Email: rapp@comp.tamu.edu}\
\vskip \baselineskip
 {\small {\it Cyclotron Institute and Physics Department, 
      Texas A\&M University, College Station, TX 77843-3366, USA}}

\vskip 2\baselineskip

\end{center}

\vskip \baselineskip

\begin{abstract}
We discuss the status of calculating in-medium modifications of
vector-meson spectral functions in hot and dense matter, their
application to dilepton spectra in ultrarelativistic heavy-ion
collisions, and possible relations to chiral symmetry restoration.
We emphasize the importance of constraining in-medium spectral 
functions by empirical information from scattering data, QCD sum 
rules, and lattice QCD. This is a mandatory prerequisite to arrive 
at reliable predictions for low-mass dileptons in heavy-ion 
collisions.
\end{abstract}

\section{Introduction}
The description of strongly interacting matter, including  
hadronic matter and the Quark-Gluon Plasma (QGP), is a challenging 
theoretical task and requires guidance from experiment.
A key issue is the determination of the spectral properties
of the excitations of the medium.
On the one hand, this allows to identify and characterize the 
relevant degrees of freedom at given temperature ($T$) and baryon
density ($\varrho_B$).
On the other hand, certain quantum-number channels bear rather direct
connections to order parameters of phase transitions, so that 
their modifications can serve as indicators of phase changes. 
In the laboratory, the only way of creating high energy-density matter
is to collide heavy nuclei at high energies. A large body
of hadronic observables has provided ample evidence
that the collision energy is largely converted into 
producing {\em matter}, justifying the
notion of approximate (local) equilibrium. However, spectral 
properties of the medium as encoded in (hadronic) correlation 
functions require the measurement of invariant-mass 
spectra of decay
products of strongly decaying resonances (long-lived hadrons, 
e.g., $J/\psi$ or $\pi^0$, decay long after the "freezeout"
of the interacting system). Strong decay channels (e.g., 
$\rho\to\pi\pi$ or $\Delta\to N \pi$), while abundant, are 
subject to final-state interactions which  
distort the invariant-mass information of the parent particle.
Nevertheless, $\pi^+\pi^-$ or $\pi^\pm p$ spectra may carry 
valuable information on the dilute-matter stages of a 
heavy-ion collision~\cite{res-exp,res-theo}. 
Dileptons (\eg, from direct decays of the vector mesons  
$\rho$, $\omega$ and $\phi$) undergo negligible
final-state interactions, and therefore emanate throughout the 
evolution of the hot and dense medium, rendering them, in principle, 
a direct probe of medium effects in the earlier 
phases~\cite{ceres,Arnaldi:2006jq,Rapp:1999ej}. 
Since the formation of sufficiently hot/dense matter in thermal 
equilibrium implies the presence of deconfinement and chiral symmetry 
restoration, the question is not so much {\em if} but
{\em how} these are realized and how to deduce them from experiment.


\section{Chiral Order Parameters and Vector Mesons}
\label{sec_order}
\begin{figure}[!t]
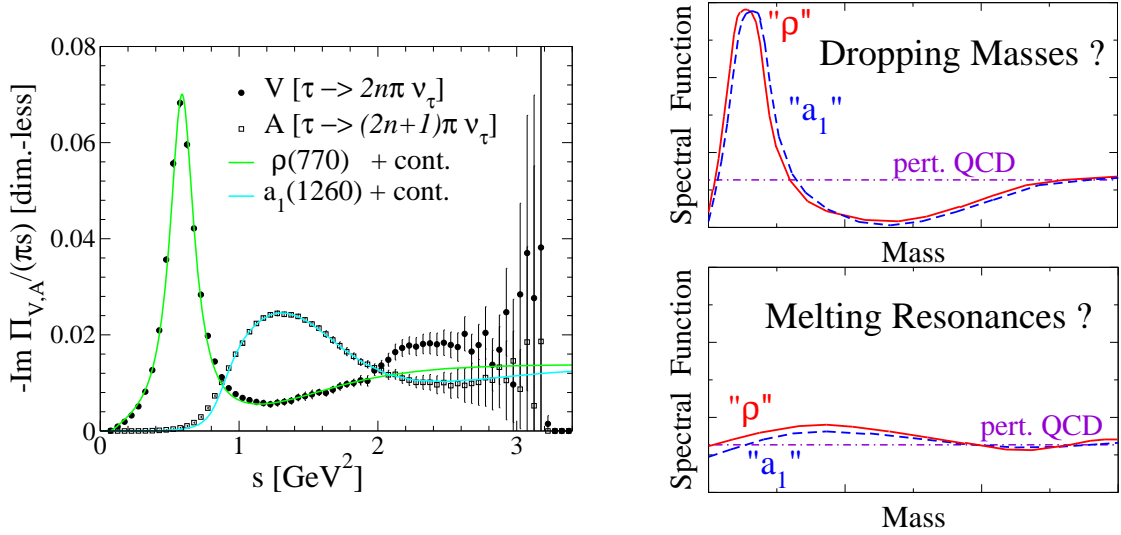

\begin{minipage}{7.5cm}
\epsfig{file=VAfit3.eps,width=7.5cm}
\end{minipage}
\hspace{1cm}
\begin{minipage}{7cm}
\epsfig{file=DM-scenario.eps,width=6cm}
\epsfig{file=RW2-scenario.eps,width=6cm}
\end{minipage}
\caption{\it Left panel: vector and axialvector spectral functions from  
hadronic $\tau$ decays~\cite{aleph98} with model fits using vacuum 
$\rho$ and $a_1$ spectral functions plus perturbative 
continua~\cite{Rapp02}; right panel: schematic scenarios for chiral 
symmetry restoration in hot and dense matter.}
\label{fig_VA}
\end{figure}

In the QCD vacuum, the spontaneous breaking of chiral symmetry (SBCS) 
is characterized by nonzero order parameters, \eg, the scalar quark 
condensate, $\langle \bar qq\rangle$$\simeq$($-250\,$MeV)$^3$, the 
pion-decay constant, $f_\pi$=93\,MeV, and the constituent quark mass, 
$m_{u,d}^\star$$\simeq$350\,MeV. Relations to hadron masses are less 
direct, but SBCS implies a massive splitting of chiral partners in 
the low-energy hadronic spectrum ($\Delta M$$\simeq$0.5\,GeV). One of 
the cleanest examples are the isovector-vector ($V$) and -axialvector 
($A$) spectral functions as measured in hadronic $\tau$ decays, 
cf.~left panel of Fig.~\ref{fig_VA}~\cite{aleph98}. 
Their integrated difference
directly relates to order parameters via chiral (or Weinberg)
sum rules~\cite{Wein67},
\begin{eqnarray}
f_n = - \int\limits_0^\infty \frac{ds}{\pi} \ s^n \ 
 \left[{\rm Im} \Pi_V(s) - {\rm Im} \Pi_A(s) \right]  \ ,  
\qquad \qquad \qquad
\label{csr}
\\ 
f_{-2} = f_\pi^2  \frac{\langle r_\pi^2 \rangle}{3} - F_A \ , \quad 
f_{-1} = f_\pi^2 \ , \quad
 f_0   = 0 \ ,  \quad
f_1 = -2\pi \alpha_s \langle {\cal O} \rangle  \ 
\label{fn}
\end{eqnarray}
($r_\pi$: pion charge radius, $F_A$: axialvector formfactor in the 
radiative pion decay, $\pi\to l \bar\nu_l\gamma$, 
$\langle {\cal O}\rangle$$\simeq$(16/9)$\langle\bar qq \rangle^2$:
4-quark condensate in factorization approximation).
As discussed below (cf.~Eqs.~(\ref{Rll}) and (\ref{Piem})), the rate
of thermal dilepton emission at low mass is dominated by the $\rho$ 
meson, providing a direct window on its in-medium spectral function. 
When combined with an evaluation of the $a_1$ spectral function within
a chiral approach, the Weinberg sum rules (which remain valid in the 
medium~\cite{KS94}) are a valuable tool to infer mechanisms of chiral 
restoration, especially when augmented with finite-$T$  
lattice QCD (lQCD) results for the $f_n$ coefficients. 
Useful insights may also be obtained from recent lQCD calculations of 
scalar and isovector quark-number susceptibilities~\cite{Allton:2005gk};
we will return to this issue at the end of  Sec.~\ref{sec_dilep}. 

\section{Vector Mesons in Medium}
\label{sec_med}
\begin{figure}[!t]
\begin{center}
\epsfig{file=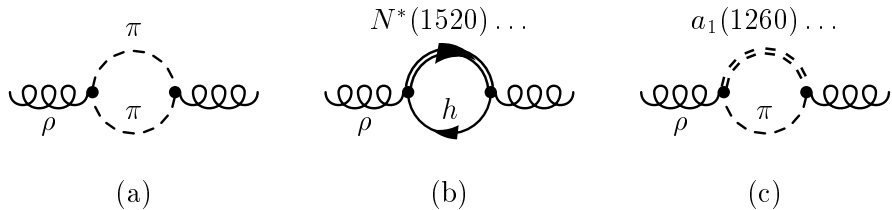,width=12cm}
\end{center}
\vspace{-0.5cm}
\caption{\it Sources of medium effects induced by interactions of 
the $\rho$ meson in hot and dense hadronic matter:
(a) renormalization of its pion cloud due to modified pion propagators,
and direct interactions of the $\rho$ meson with (b) baryons 
and (c) mesons, typically approximated by baryon- 
and meson-resonance excitations~\cite{Urban:1999im,Rapp:1999us}.
}
\label{fig_dia}
\end{figure}

\begin{figure}[!t]
\begin{minipage}{5cm}
\epsfig{file=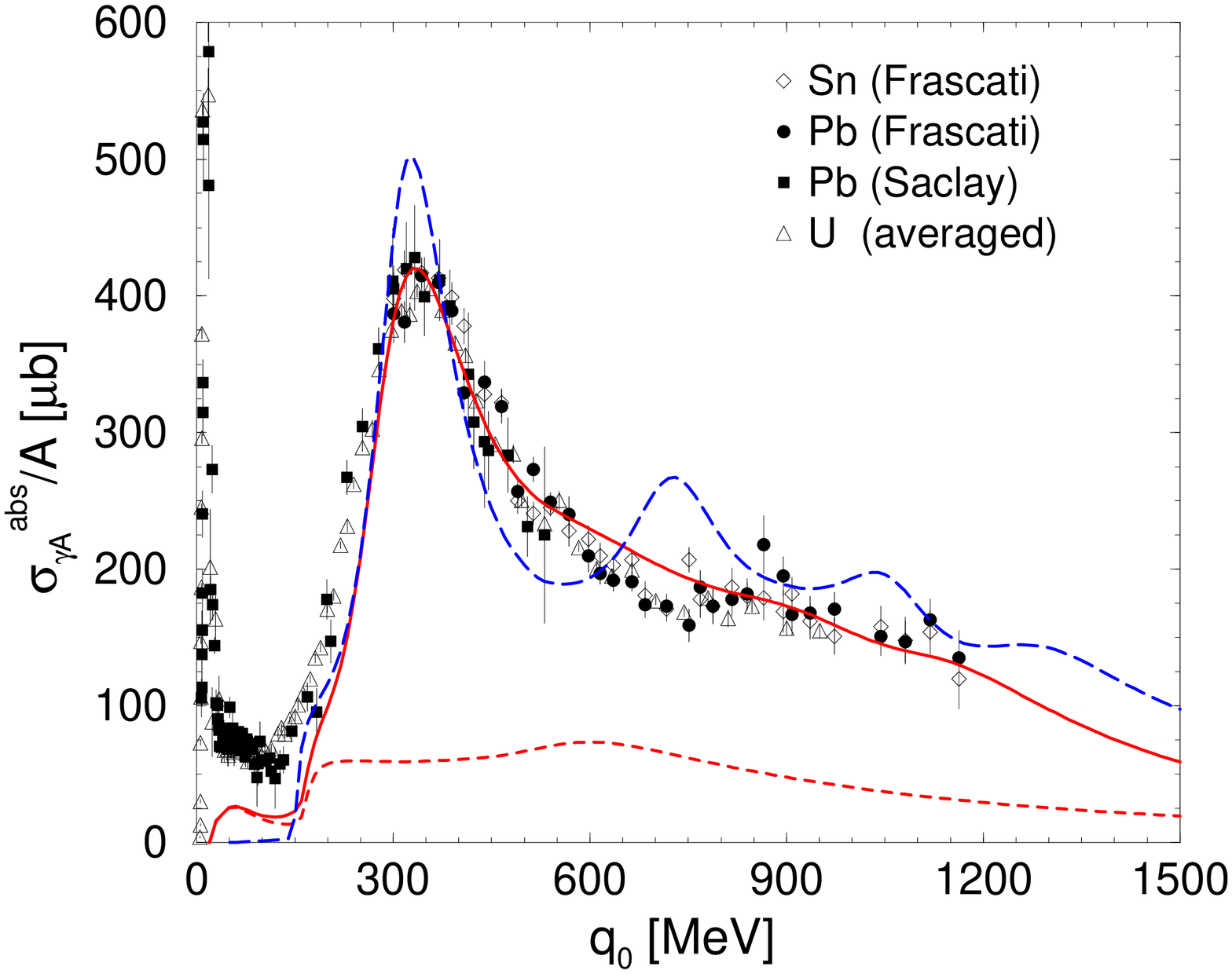,width=1.3\linewidth,angle=-90}
\hspace{1.3cm}
\epsfig{file=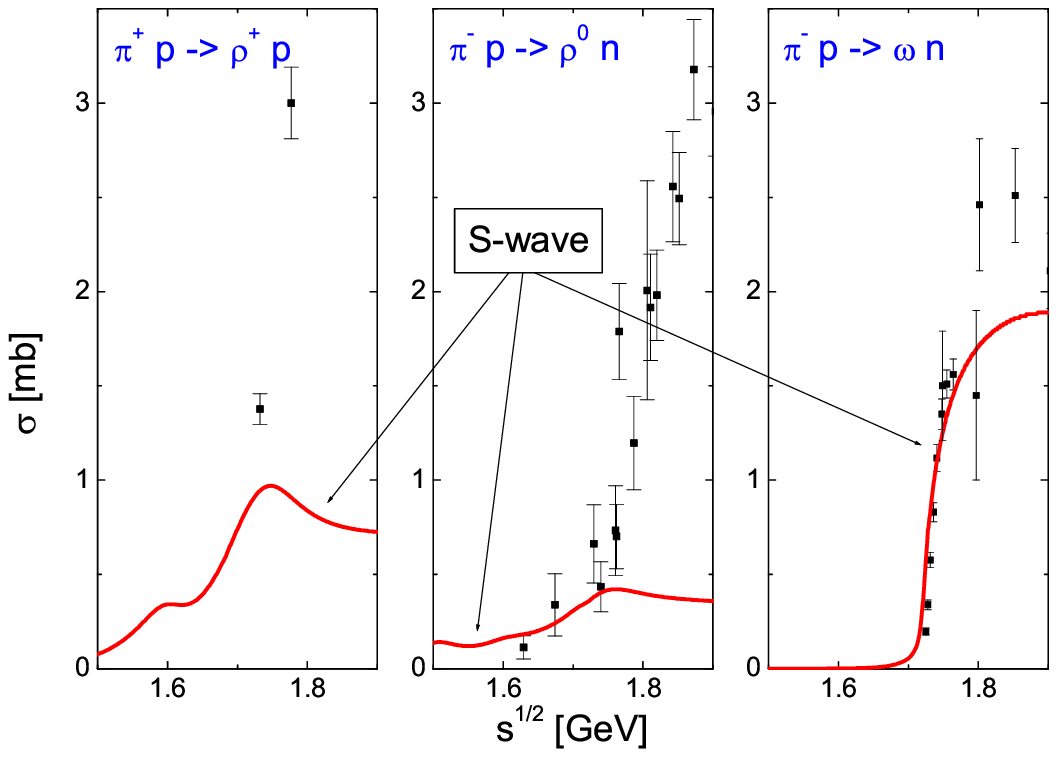,width=1.3\linewidth,angle=0}
\end{minipage}
\hspace*{4cm}
\begin{minipage}{5cm}
\vspace{0cm}
\epsfig{file=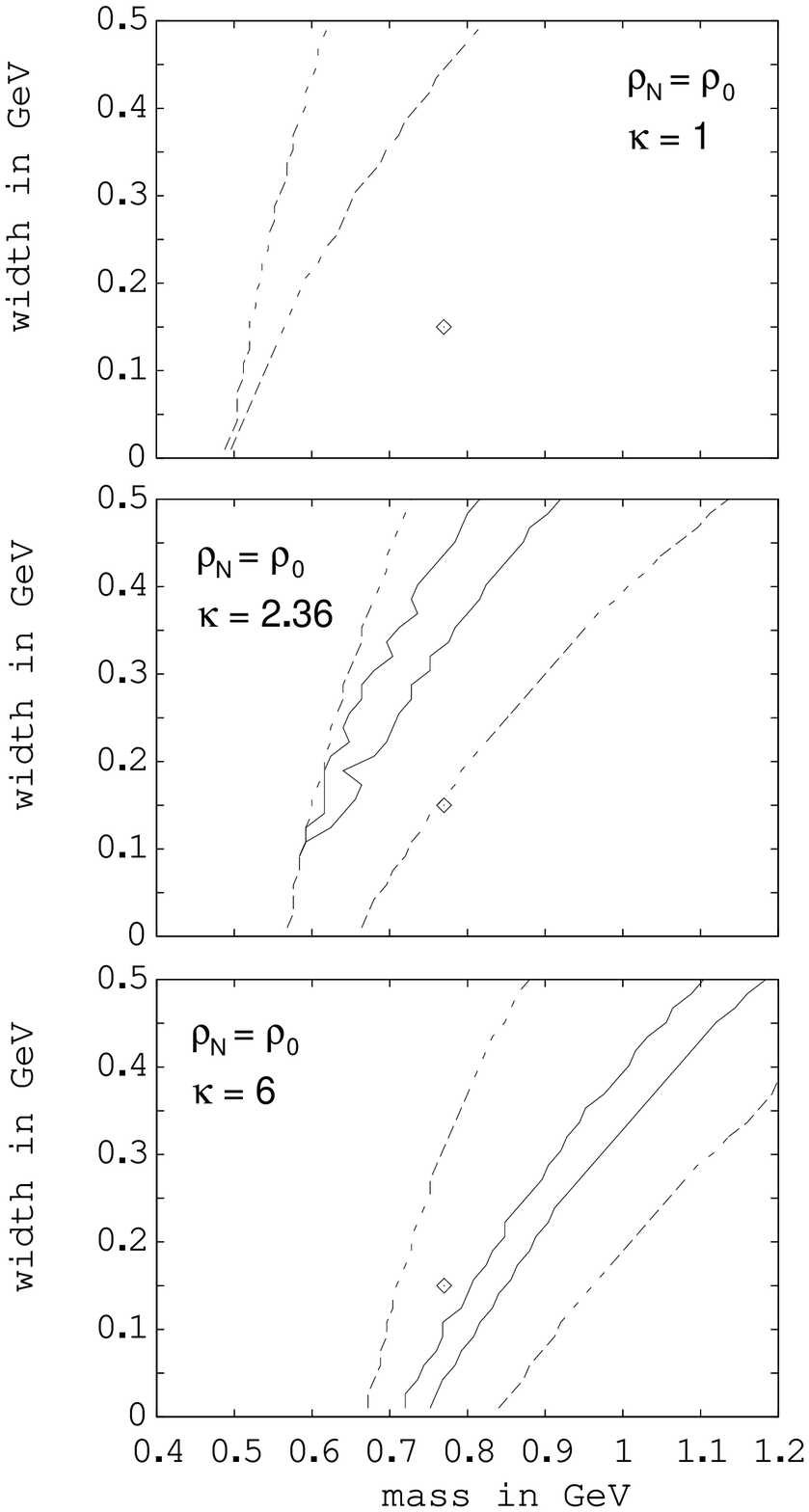,width=1.25\linewidth,angle=0}
\end{minipage}
\caption{\it Constraints on in-medium $\rho$ mesons from photoabsorption
on nuclei (upper left panel)~\cite{Rapp:1997ei}, $\pi N\to \rho N$
scattering (lower left panel)~\cite{Lutz:2001mi}, and 
QCD sum rules (right panels)~\cite{Leupold:1997dg}.}
\label{fig_constraint}
\end{figure}

The most common approach~\cite{Rapp:1999ej} to evaluate medium 
effects on vector-meson properties consists of evaluating 
effective interactions with surrounding hadrons 
from the heat bath, leading to in-medium selfenergy insertions 
which are resummed in the propagator as (here for the $\rho$)  
\be
D_\rho(M,q;\mu_B,T)= \left[M^2-(m_\rho^{(0)})^2-\Sigma_{\rho\pi\pi}
-\Sigma_{\rho B} -\Sigma_{\rho M} \right]^{-1} \ .
\ee
The different contributions may be classified as being due to
(a) modifications of the pions in the $\pi\pi$ decay, 
(b) direct $\rho$-baryon couplings (\eg, $\rho$-$N(1520)N^{-1}$
excitation) and (c) direct $\rho$-meson couplings (\eg, 
$\rho+\pi\to a_1$ excitations), cf. Fig.~\ref{fig_dia}.
The key to a reliable description of the in-medium spectral function
lies in constraining the effective vertices by independent empirical
information. Besides hadronic and radiative branching ratios
(\eg, $N(1520)\to \rho N, \gamma N$), more comprehensive information
is encoded in scattering data, \eg, $\pi N\to\rho N$ (lower left panel
in Fig.~\ref{fig_constraint}~\cite{Lutz:2001mi}), $\gamma N$ absorption 
(both representing leading order in nuclear density, $\varrho_N$), 
or $\gamma A$ absorption (upper left panel in 
Fig.~\ref{fig_constraint}~\cite{Rapp:1997ei}).
The latter is directly proportional to the $\rho$ spectral function
at {\em finite} density, 
$\sigma_{\gamma A}^{\rm abs}(q_0)\propto 
{\rm Im} D_\rho(M=0,q;\varrho_N)$.
In addition, QCD sum rules provide further checks by relating 
in-medium quark- and gluon condensates to dispersion integrals over 
the calculated spectral functions, cf.~right panel of 
Fig.~\ref{fig_constraint}~\cite{Leupold:1997dg}.

\begin{figure}[!t]
\begin{minipage}{6cm}
\vspace{-0.6cm}
\epsfig{file=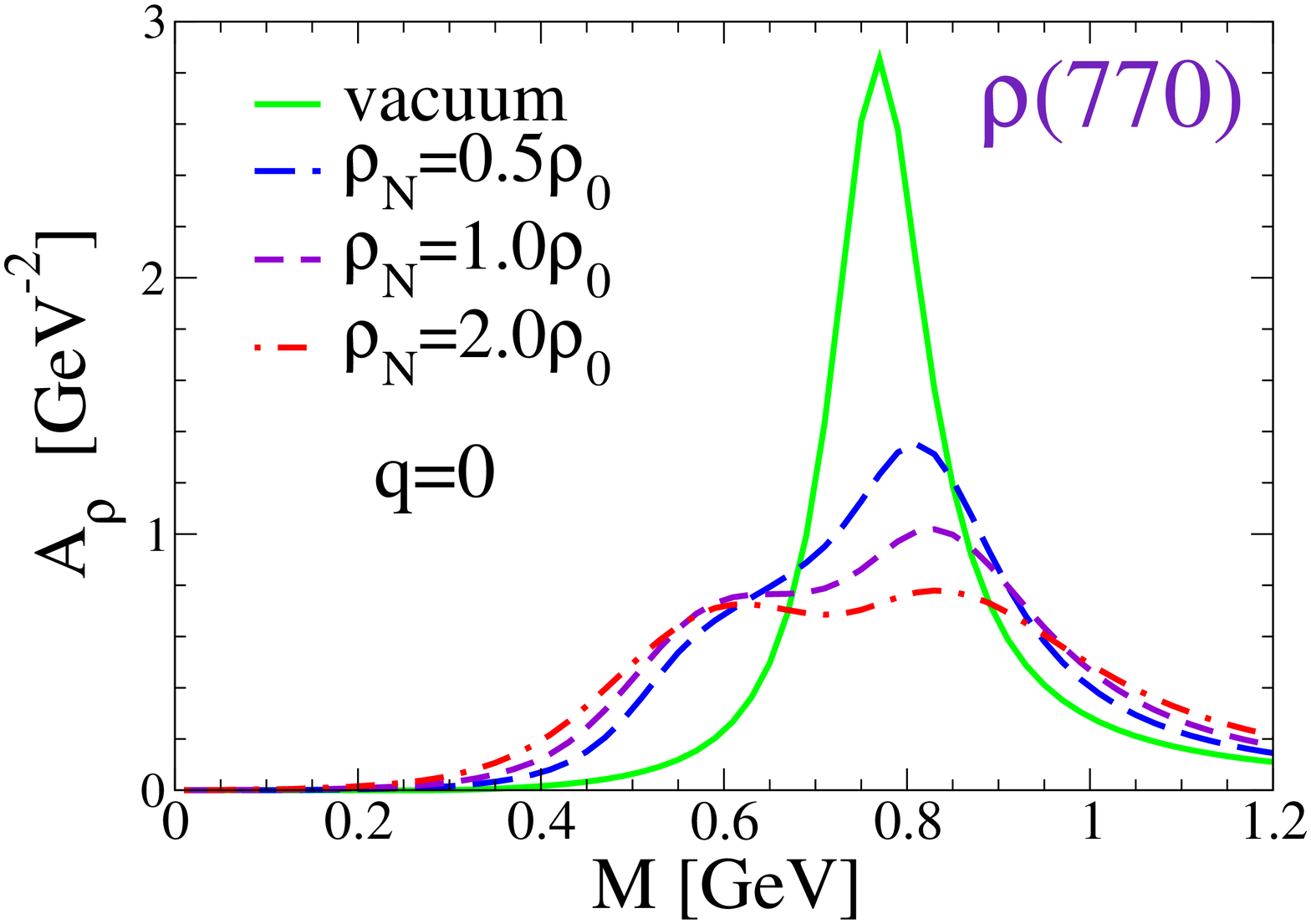,width=1.03\linewidth,angle=-90}
\end{minipage}
\hspace{1.5cm}
\begin{minipage}{6cm}
\epsfig{file=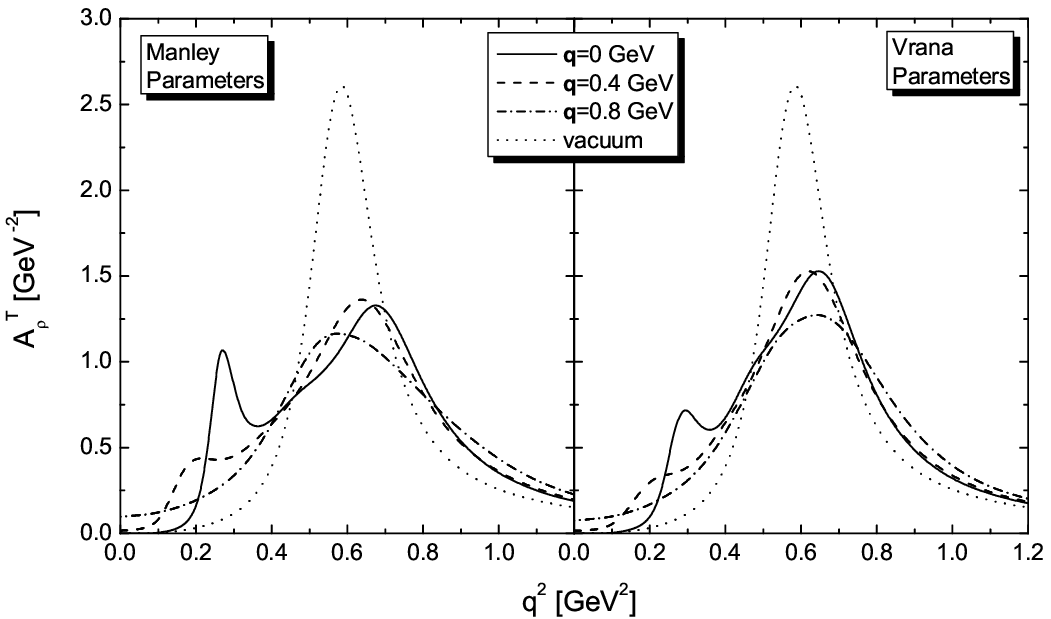,width=1.5\linewidth,angle=0}
\end{minipage}

\hspace{0.2cm}
\begin{minipage}{6cm}
\vspace{0.3cm}
\epsfig{file=Arhoc-Oset.eps,width=1.135\linewidth,angle=0}
\end{minipage}
\hspace{2cm}
\begin{minipage}{6cm}
\epsfig{file=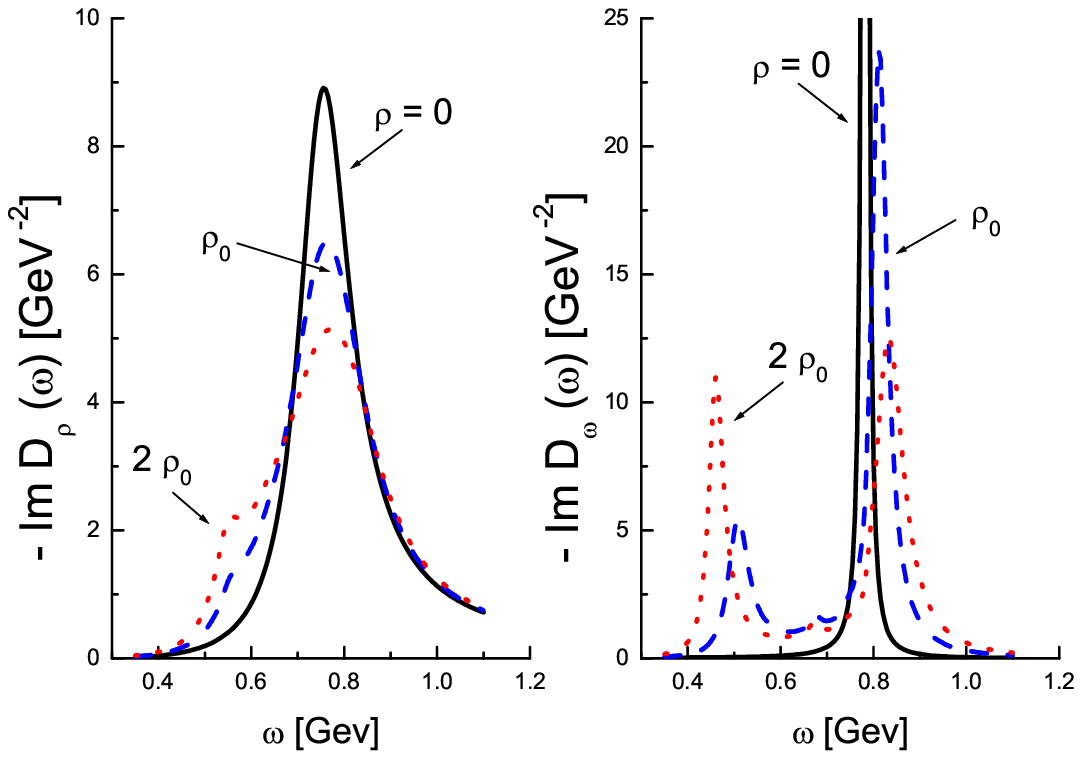,width=1.32\linewidth,angle=0}
\end{minipage}
\caption{\it Comparison of $\rho$-meson spectral functions in cold 
nuclear matter within the many-body approaches of 
Refs.~\cite{Urban:1999im,Rapp:1999us} 
(upper left panel), \cite{Post:2000qi} (upper
right panel), \cite{Cabrera:2000dx} (lower left panel) and 
\cite{Lutz:2001mi} (lower right panel). 
}
\label{fig_Arho-cold}
\end{figure}

A selection of $\rho$ spectral functions in cold nuclear
matter is compiled in Fig.~\ref{fig_Arho-cold}. 
All calculations qualitatively agree in that the main effect is a 
substantial broadening, accompanied by a small upward mass shift. 
The largest effects are found in the calculation of 
Ref.~\cite{Urban:1999im} (upper left panel), which includes an 
in-medium pion cloud and direct $\rho$-$N$ interactions and has 
been constrained by both nuclear photoabsorption and $\pi N\to\rho N$
scattering. The results agree within 30\% with a calculation where
the $\rho$-$N$ interactions are based on comprehensive
$\pi$-$N$ phase shift analyses~\cite{Post:2000qi} (upper right panel). 
This agreement is not surprising to the extent that the leading-density 
term is constrained by the same empirical data, and that higher orders 
are suppressed. The calculations of Ref.~\cite{Cabrera:2000dx} 
(lower left panel) include the dressing of the pion cloud but only
the $N(1520)$ resonance for the direct $\rho$-$N$ coupling;
consequently, the medium effects are somewhat less pronounced. This
also applies to Ref.~\cite{Lutz:2001mi} (lower right panel), where
the focus was on $S$-wave $\rho$-$N$ scattering (\ie, no $P$-wave
resonances included). 
It is noteworthy that the predicted width and mass estimates of 
Refs.~\cite{Urban:1999im,Rapp:1999us}, 
$(m_\rho,\Gamma_\rho)$$\simeq$(0.8,0.45)~GeV at nuclear saturation
density, $\varrho_N$=0.16\,fm$^{-3}$ (upper left panel in 
Fig.~\ref{fig_Arho-cold}), are in good agreement with the "allowed" 
range inferred from QCD sum rules (cf.~right middle panel in 
Fig.~\ref{fig_constraint})~\cite{Leupold:1997dg}. 

In Ref.~\cite{SYZ97} the in-medium electromagnetic (e.m.) correlation
function, $\Pi_{\rm em}$ (which governs the thermal dilepton rate), 
has been evaluated using a virial expansion in pion- and 
nucleon-densities coupled with chiral reduction formulae.  
In the soft-pion limit in a pion gas, the isovector channel exhibits
the so-called chiral mixing effect~\cite{DEI90},
\be
\Pi_{V,A} = (1-\varepsilon)~\Pi_{V,A} - \varepsilon~\Pi_{A,V} \ ,
\quad \varepsilon=T^2/6f_\pi^2 \ , 
\label{VAmix}
\ee
indicating that the $\rho$-resonance peak is not broadened albeit 
quenched. More elaborate calcu\-la\-tions lead to substantial 
strength below the $\rho$-mass, with sizable contributions from 
baryons~\cite{SYZ97}. Different 
approaches~\cite{Rapp:1999us,SYZ97,Eletsky:2001bb} thus agree 
that medium
modifications due to nucleons (baryon density, $\varrho_B$) are 
more important than due to pions (temperature, $T$), at comparable 
density. Pions are,
after all, Goldstone bosons implying reduced interaction strength.  


\section{Dilepton Spectra in Heavy-Ion Collisions}
\label{sec_dilep}
The thermal emission rate of dileptons per unit 4-volume and 
4-momentum is given by~\cite{MT85} 
\be
\frac{dN_{ll}}{d^4xd^4q} = -\frac{\alpha_{\rm em}^2}{\pi^3 M^2} \
       f^B(q_0;T) \  {\rm Im}\Pi_{\rm em}(M,q;\mu_B,T) \ .
\label{Rll}
\ee
At low mass, $M$$\le$1~GeV, where radiation from hadronic matter 
prevails, the (free)  e.m.~spectral function is saturated by the 
light vector mesons $\rho$, $\omega$ and $\phi$ (vector dominance 
model, VDM), 
\be
{\rm Im}\Pi_{\rm em} = C_\rho~{\rm Im} D_\rho +  
C_\omega{\rm Im}~D_\omega + C_\phi~{\rm Im} D_\phi \  .   
\label{Piem}
\ee
In phenomenological fits, the prefactors $C_V=m_V^4/g_V^2$ approximately 
reproduce the e.m.~decay branchings in the naive quark model, \ie,  
$C_\rho : C_\omega : C_\phi = 9:1:2$, illustrating the prevalent role of
the $\rho$-meson. 
In most approaches, (a form of) VDM is assumed to hold in the 
medium, although this has been challenged recently~\cite{Harada:2003jx}.  

\begin{figure}
\begin{minipage}[t]{6cm}
\epsfig{file=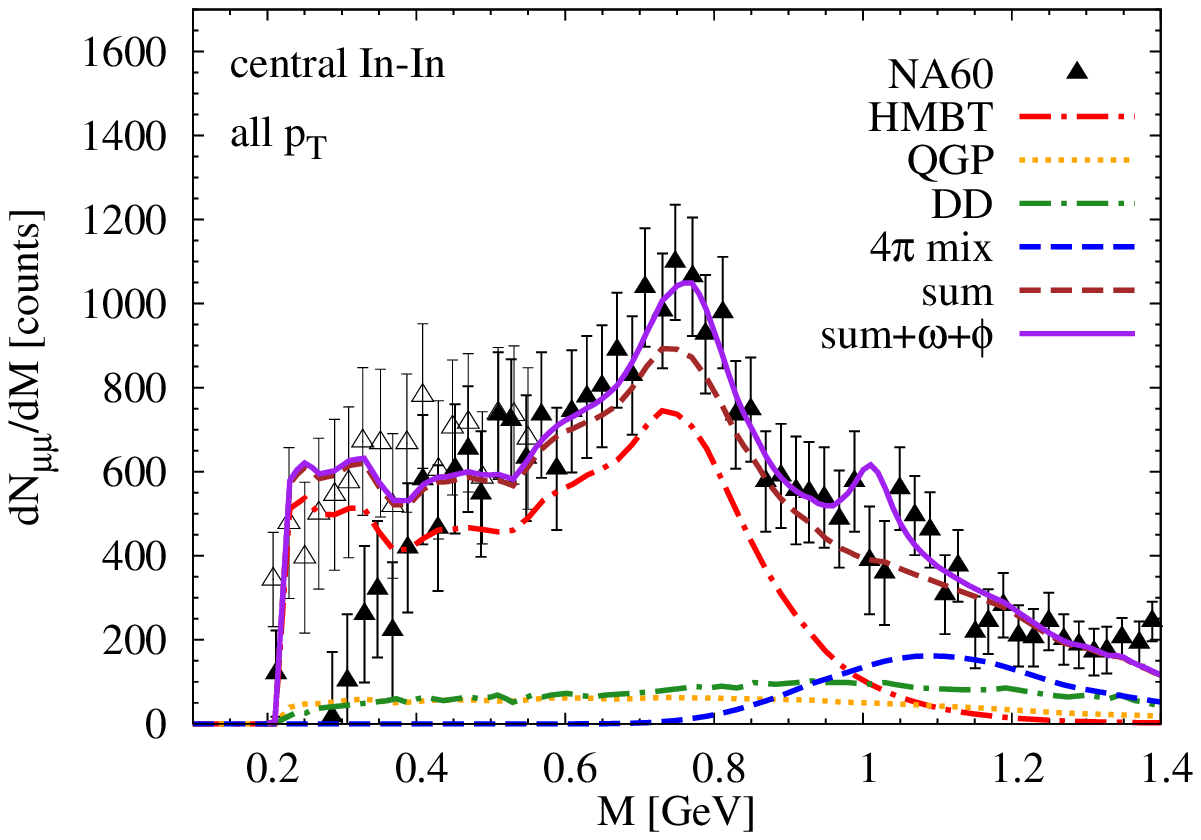,width=1.38\linewidth,angle=0}
\end{minipage}
\hspace{2.3cm}
\begin{minipage}[t]{6cm}
\epsfig{file=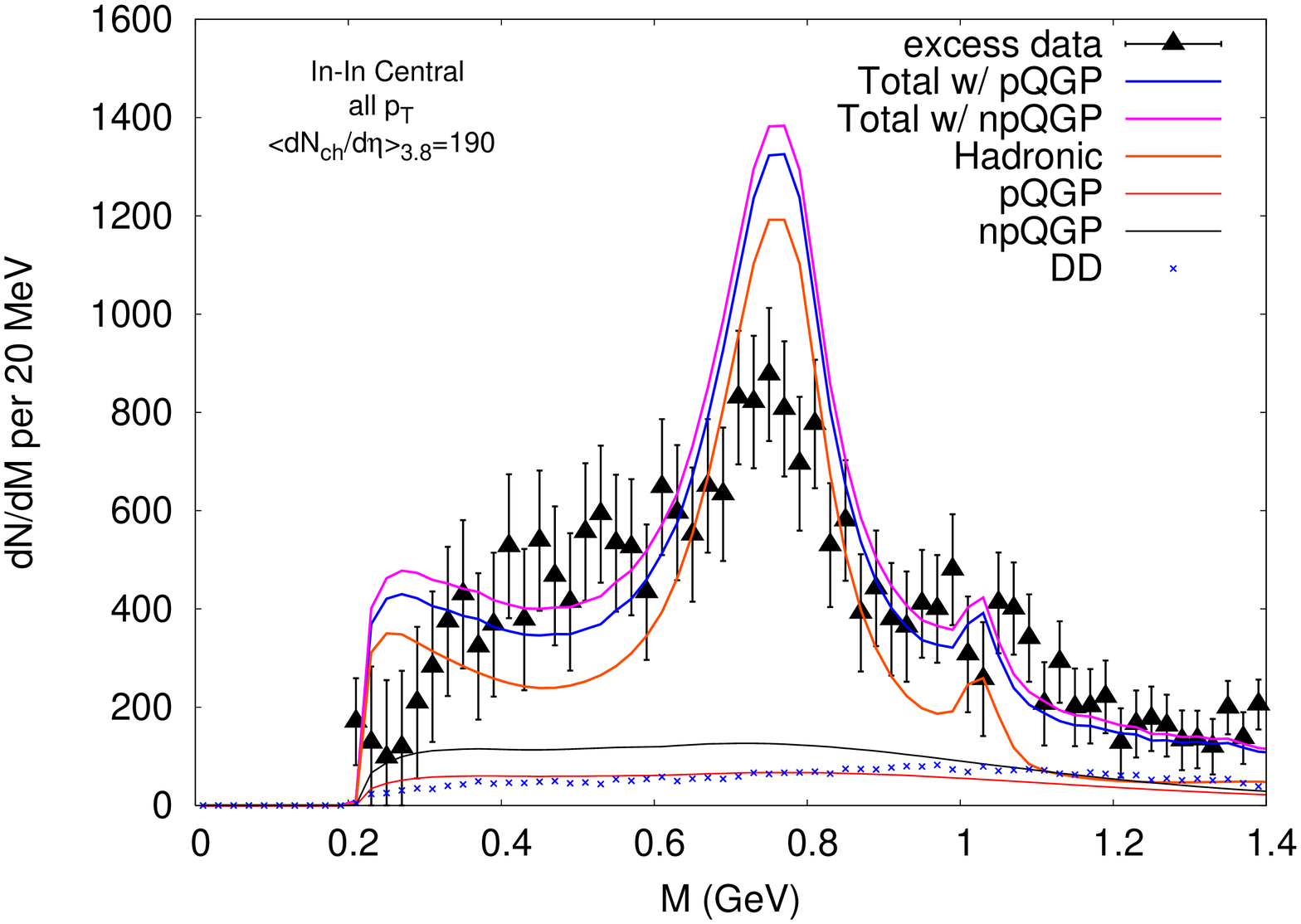,width=1.42\linewidth,angle=0}
\end{minipage}
\caption{\it NA60 dilepton excess spectra in central 
$In$-$In$~\cite{Arnaldi:2006jq} compared 
to calculations using (a) vector-meson spectral functions from hadronic 
many-body theory~\cite{Rapp:1999us} within an expanding fireball 
(left)~\cite{vanHees:2006ng}, (b) a chiral virial expansion within 
a hydrodynamic simulation (right)~\cite{Dusling:2006yv}.}
\label{fig_na60}
\end{figure}

The $T$- and $\varrho_B$-dependent emission rates are to be  
convoluted over the space-time history of a heavy-ion reaction. In the 
left panel of Fig.~\ref{fig_na60}, predictions based on an in-medium 
$\rho$ spectral function, coupled with a thermal fireball model for
central $In$(158\,AGeV)-$In$ collisions at SPS, are compared to new NA60 
data~\cite{Arnaldi:2006jq}. 
The latter have reached a level of precision that allows to subtract 
known background sources to
isolate the dilepton {\em excess} spectra. The calculations,
supplemented with (small) contributions from in-medium $\omega$ and
$\phi$ decays~\cite{Rapp:2000pe} as well as QGP emission, describe the 
data well~\cite{vanHees:2006ng}.
The absolute yield carries an uncertainty due to the fireball
lifetime ($\tau_{FB}$$\simeq$7\,fm/c) of about 25\%, but the shape 
and relative contributions of the spectrum are robust.
The calculations imply that the underlying $\rho$ spectral function is 
essentially "melted" in the vicinity of the critical 
temperature, $T_c$$\simeq$175\,MeV, with no significant mass shift 
at smaller temperatures. In fact, close to $T_c$, the 
pertinent e.m.~emission rate resembles the perturbative QGP rate which 
is suggestive for a form of quark-hadron duality~\cite{Rapp:1999us}. 
This conclusion is corroborated by a hydrodynamic convolution of the 
rates following from the chiral virial approach (right panel of 
Fig.~\ref{fig_na60})~\cite{Dusling:2006yv}, where the low-mass 
enhancement is accounted for but the lack of $\rho$-broadening
results in a $\sim$40\% overestimate of the data at the free $\rho$-mass.
When folding the chiral virial rates over an expanding 
fireball model~\cite{vanHees:2006iv}, both hadronic and QGP spectra 
agree well with the hydrodynamic results of Ref.~\cite{Dusling:2006yv}.
The fireball model of Ref.~\cite{Renk:2006ax} deviates from
these findings in that (i) the low-mass NA60 data are fitted with 
a $\rho$ spectral function based on pion-gas effects only~\cite{Riek06}, 
(ii) the intermediate-mass region (IMR, $M$$>$1\,GeV) is dominated by 
QGP radiation. 
\begin{figure}[!t]
\vspace{-0.7cm}
\begin{minipage}[t]{6cm}
\hspace{1cm}
\epsfig{file=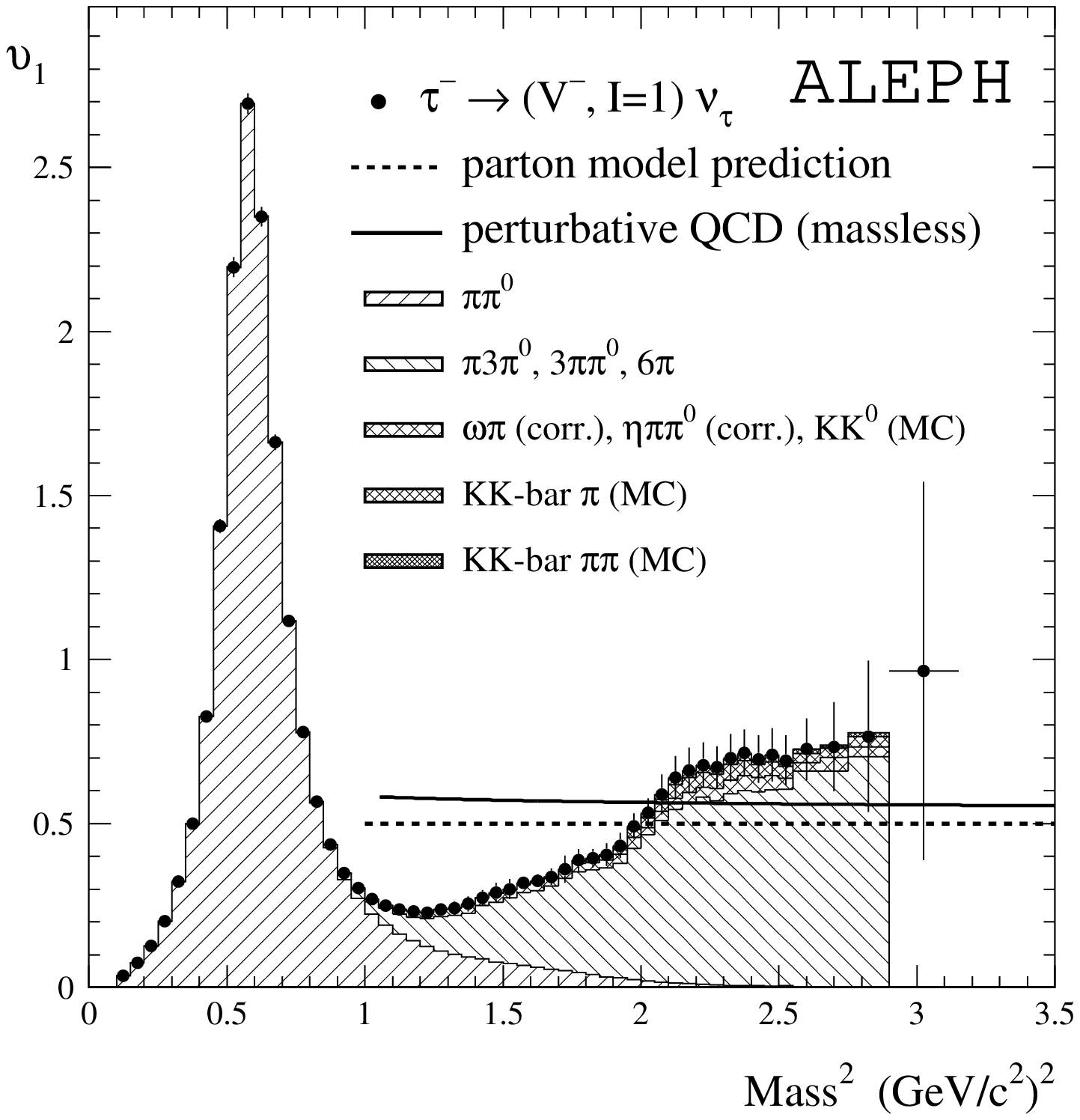,width=1.6\linewidth,height=1.1\linewidth,
angle=0}
\end{minipage}
\hspace{2cm}
\begin{minipage}[t]{6cm}
\epsfig{file=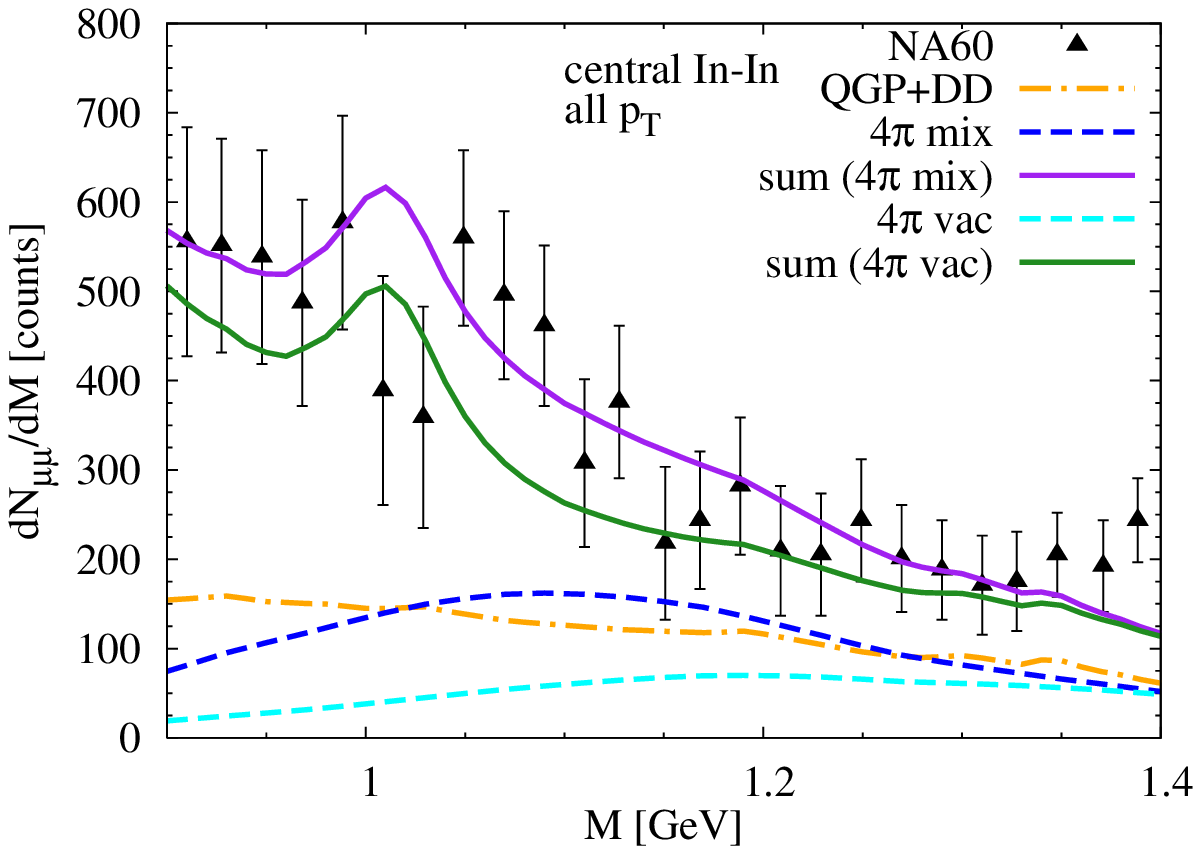,width=1.4\linewidth,angle=0}
\end{minipage}
\caption{\it Left panel: decomposition of the free vector-isovector
spectral function into 2$\pi$ and 4$\pi$ components as measured
in $\tau$ decays~\cite{aleph98}; right panel: 4$\pi$ contributions
to intermediate-mass dileptons in NA60~\cite{Arnaldi:2006jq} with/-out
(upper/lower dashed line) medium effects~\cite{vanHees:2006ng}.}
\label{fig_imr}
\end{figure}
In the approaches underlying Fig.~\ref{fig_na60}, 4-pion type 
annihilations (\eg, $\pi$+$a_1$$\to$$\mu^+\mu^-$), which follow from 
the e.m.~correlator in free space (cf.~left panel of 
Fig.~\ref{fig_imr}), significantly feed into the IMR. 
Lower and upper estimates based on the free 4-$\pi$ part of the 
e.m.~spectral function and on the naive chiral mixing formula
(Eq.~(\ref{VAmix}) assuming $\varepsilon(T_c)$$\equiv$1/2),
provide a band consistent with the NA60 data (upper solid curves 
in Fig.~\ref{fig_imr}, right panel)~\cite{vanHees:2006ng};   
the chiral virial approach is consistent with this band 
(Fig.~\ref{fig_na60}, right panel)~\cite{Dusling:2006yv}.   

The NA60 data are not yet sensitive to extract information on 
medium modifications of $\omega$ and $\phi$ mesons~\cite{Rapp:2000pe} 
as included in the left panel of Fig.~\ref{fig_na60}. 
It is interesting to note that a recent
measurement of photon-induced $\omega$-production off nuclei in the
$\pi^0\gamma$ decay channel~\cite{CBTAPS} is indicative for an 
in-medium reduced $\omega$-mass, together with significant broadening.    
Moreover, 2-flavor finite-$T$ lQCD calculations find that when
approaching the (putative) critical point by increasing the 
quark-chemical potential,
the isoscalar susceptibility ($\omega$ channel) develops a rather sharp
maximum while the isovector one ($\rho$ channel) behaves 
smoothly~\cite{Allton:2005gk}.
The latter is consistent with hadronic many-body calculations
of the $\rho$ meson which exhibit strong broadening without
significant mass shift under both RHIC and SPS conditions. 
At the same time, the effect of $\sigma$-$\omega$ mixing, together 
with a softening of a $\sigma$ mode at the critical point, could
induce a dropping-mass component in the $\omega$ spectral function. 

\section{Conclusions}
\label{sec_concl}
The study of in-medium vector mesons provides rich
information on spectral properties of QCD matter including its
phase structure. "Quality control" in hadronic model approaches
by imposing empirical and theoretical constraints is mandatory 
to allow for meaningful applications to ultrarelativistic heavy-ion
collisions. Recent high-precision dilepton data 
at SPS energies are consistent with
hadronic many-body calculations which predict a "melting" of the 
$\rho$ resonance close to the expected phase boundary. The full
potential of the data, including transverse-momentum and centrality
dependencies, has yet to be exploited. 
Connections to chiral symmetry restoration through the axialvector
spectral function should be pursued with high priority, both 
theoretically and experimentally ($\pi^\pm \gamma$ 
spectra?). Future measurements at low and high {\em net} baryon
density at RHIC and the Compressed
Baryonic Matter experiment at GSI will be essential to enlarge 
the scope across the QCD phase diagram.

\vskip\baselineskip

{\bf Acknowledgments.}
I thank H.~van Hees for his important contributions to the presented
subjects, and J.~Wambach for discussion.
This work was supported in part by a U.S. National Science Foundation
CAREER Award under grant PHY-0449489.


\end{document}